\documentclass[aps,prb,amsmath,amssymb,showpacs,superscriptaddress,twocolumn,floatfix]{revtex4-1}

\usepackage{amssymb}
\usepackage{graphicx}
\usepackage{enumerate}
\usepackage{epsfig,amsmath}
\usepackage{subfigure}
\usepackage{hyperref}

\usepackage{color}
\usepackage[usenames,dvipsnames]{xcolor}
\usepackage{multirow}
\usepackage{braket}

\begin{document}
\title{Hanbury--Brown and Twiss exchange effects in a four-terminal tunnel junction}

%
%
\author{Jayanta Sarkar}
\affiliation{Low Temperature Laboratory,  O.V. Lounasmaa Laboratory, Aalto University, P.O. Box 15100, FI-00076 AALTO, Finland}
\author{Ciprian Padurariu}
\affiliation{Low Temperature Laboratory,  O.V. Lounasmaa Laboratory, Aalto University, P.O. Box 15100, FI-00076 AALTO, Finland}
\author{Antti Puska}
\affiliation{Low Temperature Laboratory,  O.V. Lounasmaa Laboratory, Aalto University, P.O. Box 15100, FI-00076 AALTO, Finland}
\author{Dmitry Golubev}
\affiliation{Low Temperature Laboratory,  O.V. Lounasmaa Laboratory, Aalto University, P.O. Box 15100, FI-00076 AALTO, Finland}
\author{Pertti J. Hakonen}
\affiliation{Low Temperature Laboratory,  O.V. Lounasmaa Laboratory, Aalto University, P.O. Box 15100, FI-00076 AALTO, Finland}

\begin{abstract}
We investigate the current-current correlations in a four-terminal Al-AlOx-Al tunnel junction where shot noise dominates. 
We demonstrate that cross-correlations in the presence of two biasing sources of the Hanbury--Brown and Twiss type are much stronger (approximately twice) than an incoherent sum of correlations generated by single sources.  The difference is due to voltage fluctuations of the central island that give rise to current-current correlations in the four contacts of the junction. Our measurements are in close agreement with results obtained using a simple theoretical model based on the theory of shot noise in multi-terminal conductors, generalized here to arbitrary contacts.
\end{abstract}


\maketitle

\section{Introduction}

Shot noise arises due to the discrete nature of elementary charge carriers \cite{M.1997, Blanter2000}. An example of a classical shot noise system is the vacuum diode where electron emission events from the cathode are uncorrelated.  The time interval between successive current pulses is random and is well described by a Poisson distribution. In general, the shot noise power is given by $S_I=F 2e \braket{I}$ where $F$ is the Fano factor and $\braket{I}$ is the average current. Uncorrelated transport is described by Fano factor $F=1$.

In mesoscopic conductors, quantum interference gives rise to finite correlations between individual transport events. These quantum effects produce non-classical shot noise that typically depends on the transparency of electron transport channels. For a conductor with $n$ channels characterized by transmission eigenvalues $T_n$, the Fano factor is given by \cite{nazarov} $F= \sum_n T_n(1-T_n)/\sum_n  T_n$. Additional correlations in nano-sized conductors are commonly due to the Coulomb interaction or the Pauli exclusion principle \cite{Buttiker1990, Jong1995, Bocquillon2013}. These correlations may give rise to the bunching\cite{Kiesslich2008} ($F>1$) or the anti-bunching\cite{Henny1999a} of electrons ($F<1$).

For a two terminal conductor the shot noise at zero-frequency is given by\cite{Blanter2000} 
\begin{align}
S=\frac{e^2}{\pi \hbar} \sum_n \int dE &\left\{T_n(E) \left[f_L(1- f_L)+f_R(1- f_R)\right] + \right. \notag\\
\ &\left. T_n(E) \left[1-T_n(E)\right] (f_L-f_R)^2\right\} , \label{shotnoise}
\end{align} 
where $f_L$ and $f_R$ denote the Fermi distribution functions of electrons in the left and the right leads, respectively. The first two terms describe noise in equilibrium, while the third term is the out-of-equilibrium contribution depending on the bias via $(f_L - f_R)^2$.

Two terminal measurements of shot noise are often insufficient to uniquely identify the details of the transport regime and distinguish between predictions of different theoretical models \cite{Langen1997,Sukhorukov1999,Gutman2001}. For this reason, four-terminal measurements in the Hanbury--Brown and Twiss (HBT) configuration \cite{HanburyBrown1956}, where current can be injected into two terminals simultaneously, have proven useful \cite{Oliver1999, Oberholzer2000, Kiesel2002, Jeltes2007}.

 In this work, we have investigated HBT correlations generated by a four-terminal junction with tunnel contacts. Weak transmission of the contacts may lead to the expectation of uncorrelated transport \cite{Kumar1996}, which is the case for the shot noise generated by each contact, $F=1$, but not for the HBT current correlations. On the contrary, HBT correlations reach close to the maximum theoretical value for our sample \cite{Langen1997}. We have compared our experimental results with calculations using a simple theoretical model where HBT correlations arise due to the fluctuations of voltage in the junction region. Our data match closely the theoretical predictions, thereby confirming the assumption that the junction region is characterized by an out-of-equilibrium distribution function.

In Sec.~\ref{sec:exp} we describe the setup and experimental techniques. Sec.~\ref{sec:HBT} outlines the theoretical model, derives the expression of HBT current-current correlations, and presents the comparison between measured data and the theoretical prediction. Sec.~\ref{sec:concl} presents our conclusions.

\section{Experimental techniques}
\label{sec:exp}

\begin{figure}[!t]
\begin{center}
\includegraphics[width=0.85\linewidth]{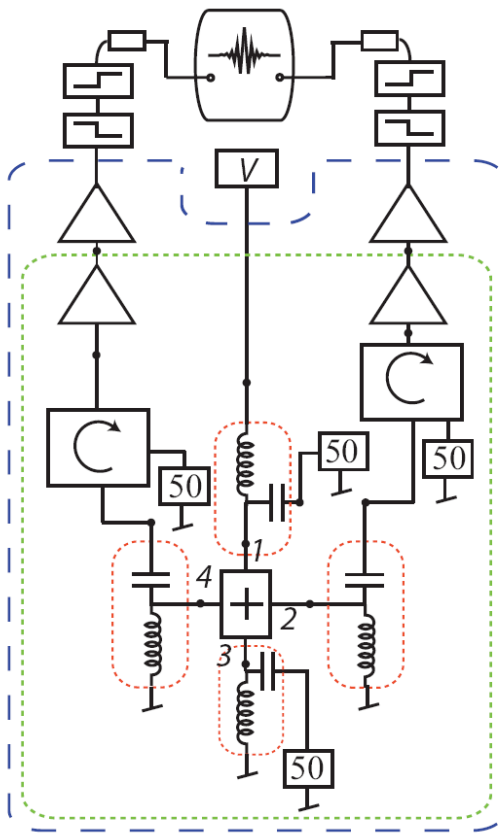}
\caption{(Color online.) Schematic of the experimental setup for  cross-correlation measurements.
The above schematic is wired for the biasing case where only terminal 1 is biased
and terminals 2-4 are grounded at low frequencies.
The cross-correlation spectrum is detected across terminals 2 and 4. There are
two circulators in the rf lines which cuts down the back action noise coming
from the cooled preamplifiers. After the amplification by room temperature amplifiers
the signals are taken through 900 MHz low pass and 600 MHz high pass filters and fed into a Le Croy
oscilloscope.}
\label{setup}
\end{center}
\end{figure}

The experiments were performed using a cross-correlation spectrometer illustrated in Fig.~\ref{setup}. In the measurement
configuration all the four terminals of the sample are connected to bias-T components, marked with red dotted lines in Fig.~\ref{setup},
allowing to apply bias to all terminals. In the present experiments, bias voltage has been applied via terminals 1 and 3 while the cross-correlation was measured between terminal 2 and 4. The cold amplifiers provided a gain of about 12 dB over the band $600 - 900$ MHz. The spectrometer had a system noise temperature of $\sim 15$ K. The circulators provided an isolation of 18 dB which was sufficient to cut down the cross talk between the channels originating from the back action noise of the preamplifiers.

The cross-correlation was calculated for two band pass filtered noise signals ($f=600 -900$ MHz) in time domain. The gains of the two channels were 104 dB and the traces were digitized using a 6 GHz Le Croy oscilloscope with (over)sampling rate of 5GS/s. The electrical length of the channels were adjusted as nearly equal which allowed us to calculate the cross-correlation as a direct array product using zero time offset in the numerical calculation.  It is known that for signals with band width $BW$, the cross-correlation is influenced by the time offset $\delta t$ only if the difference would approach $1/BW$. In our experiments, $\delta t << 1/BW$.

\section{HBT in a four-terminal tunnel junction}
\label{sec:HBT}

\subsection{Theoretical predictions}

\begin{figure}[!t]
\begin{center}
\includegraphics[width=0.47\textwidth]{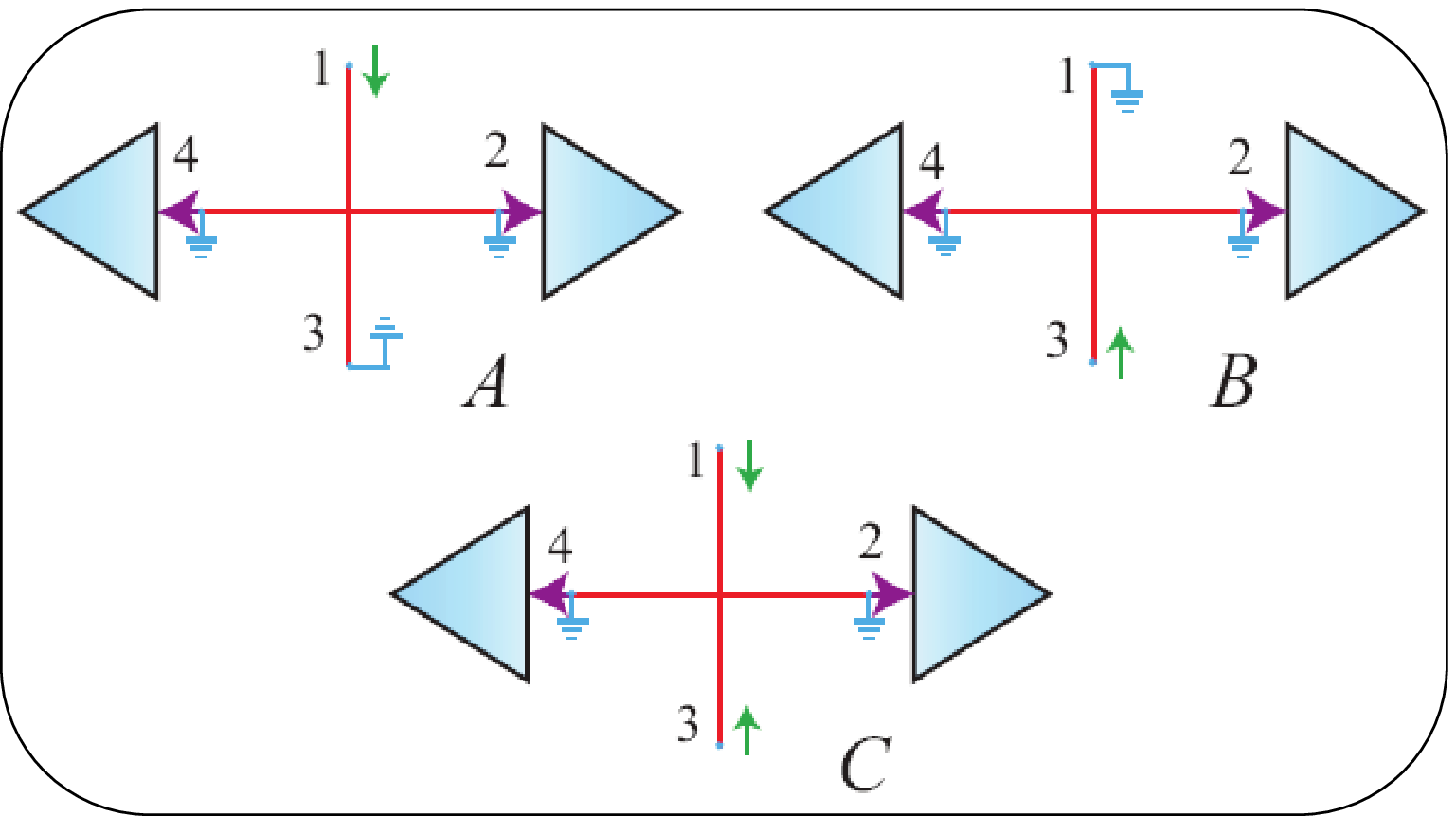}\\
\includegraphics[width=0.38\textwidth]{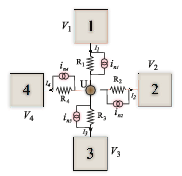}
\caption{(Color online.) Above. Definition of the three bias configurations used in the HBT experiments with $V_2=V_4=0$ and $A$) $V_1=V$, $V_3=0$; $B$) $V_1=0$, $V_3=V$; and $C$) $V_1=V_3=V$. Below. The model, similar to Ref. \onlinecite{Langen1997}, showing the equivalent circuit representation of the four-terminal junction with electric potential $U$ on the central metallic island and uncorrelated noise sources $i_{n,j}$ in each arm.}
\label{tunneljunctionlangen}
\end{center}
\end{figure}

The HBT exchange correction factor is defined in accordance to Ref. \onlinecite{M.1997} as
$ \Delta S_{nm}=|S^C_{nm}|-(|S^A_{nm}|+|S^B_{nm}|) $, where $S^A_{nm}$, $S^B_{nm}$, and $S^C_{nm}$ $(n\neq m)$ are the 
cross-correlated noise powers 
in three different configurations {\it A, B} and {\it C} as depicted in Fig. \ref{tunneljunctionlangen}.
 In a normal metallic system all cross-correlations are negative, $S^A_{nm},S^B_{nm},S^C_{nm}< 0$ for $(n\neq m)$. However, the sign and magnitude of $ \Delta S_{nm}$ will vary depending on the details of the transport \cite{Tan2016}.

It has been predicted that the HBT effect leads to large values of $\Delta S_{nm}$ for junctions in the classical shot noise regime, reaching the maximal value $(|S^A_{nm}|+|S^B_{nm}|)$. This was first considered by van Langen and B\"uttiker\cite{Langen1997} as a limiting case in their work on current correlations in a chaotic quantum dot. We have used a similar approach to generalize their result to arbitrary contacts.


The junction area is represented in our model as a metallic island with negligible level spacing connected to four macroscopic leads by contacts labeled $j=\{1,2,3,4\}$. The intrinsic resistance of the island is neglected in comparison to the resistances of the contacts. To model noise, we add to each contact a source of random current $i_{n,j}$, as in Fig.~\ref{tunneljunctionlangen}. The added noise is uncorrelated across the junction, $\langle{i_{n,j}(0)i_{n,k}(t)}\rangle=0$ for $k \neq j$. The spectral density of noise at zero frequency and for low temperature, $eV\gg k_BT$, is given by the two-terminal shot noise formula, Eq.~\ref{shotnoise}. 
\begin{align}
S_j= & \int dt \langle{i_{n,j}(0)i_{n,j}(t)}\rangle= G_j F_j \int dE (f_j-f_c)^2 +\notag\\
\    & G_j \int dE \left[f_j(1- f_j)+f_c(1- f_c)\right]. 
\label{barenoise}
\end{align} 
Here we have ignored the energy dependence of the transmission coefficients and expressed the shot noise in terms of the contact conductance, $G_j=(e^2/\pi\hbar)\sum_n T_n^{(j)}$, and the Fano factor, $F_j=\sum_n T_n^{(j)}(1-T_n^{(j)})/(\sum_n T_n^{(j)})$. The distribution function of terminal $j$ is given by, $f_j=(1+\exp[(E-eV_j)/k_BT])^{-1}$, and the out-of-equilibrium distribution function of the central metallic island is obtained by the weighted average, $f_c=\sum_j G_j f_j/G_{\Sigma}$, with $G_{\Sigma}=\sum_j G_j$. Integrating Eq.~\ref{barenoise} over energy in the limit of low temperature we find,
\begin{align}
S_j=\ & eF_jG_j\displaystyle\sum_{k=1}^{4} \frac{G_k}{G_{\Sigma}}|V_k-V_j|+  \label{S_j} \\
\ &    e(1-F_j)G_j\displaystyle\sum_{k,l=1}^4\frac{G_kG_l}{2G_{\Sigma}^2}|V_k-V_l|. \notag
\end{align} 

The measured correlations of current fluctuations $\delta I$ correspond to the total current flowing in each terminal, $S_{nm}= \int dt \langle{\delta I_{n}(0)\delta I_{m}(t)}\rangle$. At low frequencies, charging dynamics of the junction capacitance play no role. The fluctuations of the total current are solely due to the added random currents $i_{n,j}$ and the fluctuations $\delta U$ of the electric potential of the metallic island, 
\begin{align}
\delta I_{j} = -G_j \delta U + i_{n,j}, \quad \delta U = G_{\Sigma}^{-1}\displaystyle\sum_{j=1}^{4} i_{n,j},
\end{align}
where the potential fluctuations $\delta U$ are obtained from imposing current conservation, $\sum_j\delta I_{j}=0$.
Current-current correlations can now be expressed in terms of the uncorrelated shot noise contributions,
\begin{align}
S_{kl} = S_k \delta_{kl} - \frac{(G_kS_l+G_lS_k)}{G_{\Sigma}} + \frac{G_kG_l}{G^2_{\Sigma}}\displaystyle\sum_{j=1}^{4} S_j.
\label{totalnoise}
\end{align}

In the limit of identical tunnel contacts $G_j=G$ and Poisson shot noise $F_j=1$, we recover the noise correlations in a classical circuit of resistors \cite{Langen1997},
\begin{equation}
S_{kl}=\frac{e G}{8} \sum_{m=1}^4 [1+4(2 \delta_{kl}-1)(\delta_{mk}+\delta_{ml})] |V_m-U|,
\label{langeneqn}
\end{equation}
with average value of the metallic island potential $U=\sum_j V_j/4$.

In the limit of contacts formed by diffusive wires where transport is characterized by Fano factors $F_j=1/3$, Eq.~\ref{totalnoise} recovers the result obtained in Ref.~\onlinecite{Sukhorukov1999}, given below for the case of interest $k\neq l$,
\begin{equation}
S_{kl}=-\frac{e}{3}\frac{G_kG_l}{G^2_{\Sigma}} \sum_{m=1}^4 G_m \left(|V_m-V_k|+|V_m-V_l|\right).
\label{sukhorukoveqn}
\end{equation}

Throughout this paper we discuss cross-correlations between terminals $k=2$ and $l=4$, and hence for convenience, we omit the subscript and denote $-S^{\sigma}_{24}$ by $S^{\sigma} (\sigma=A,B,C)$ for biasing configurations $A$, $B$ and $C$, respectively. Since in our junction the cross-correlations are negative, the minus sign ensures that $S^{\sigma}$ is positive. The HBT exchange correction factor is given by $\Delta S = S^C-(S^A+S^B)$. We have used Eq.~\ref{totalnoise} to calculate the current-current correlations and compare with the measurements.

\subsection{Experimental results}

\begin{figure}[!t]
\begin{center}
\includegraphics[width=0.85\linewidth]{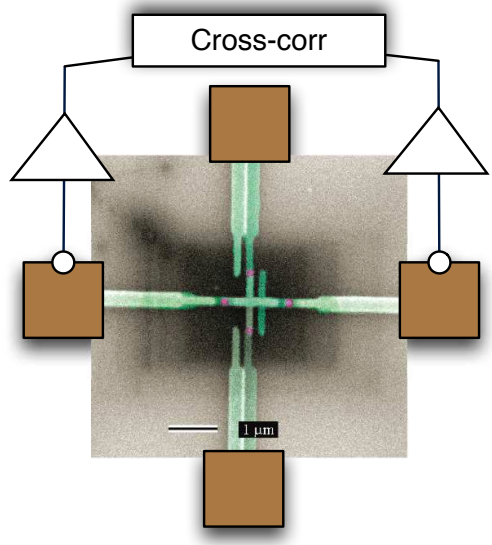}
\caption{(Color online.) Scanning electrom micrograph image of the four-terminal tunnel junction and its connection to the cross-correlation measurement system. The purple areas indicate the tunnel junctions with area $100 \times 100$ nm$^2$.}
\label{tunneljunctionexch}
\end{center}
\end{figure}

A scanning electron microscope image of our four-terminal tunnel junction is displayed in Fig. \ref{tunneljunctionexch} together with its connection to the cross-correlation measurement system; a reminder of the experimental configurations $A$, $B$, and $C$ is displayed in Fig. \ref{tunneljunctionlangen}. The sample was fabricated on sapphire wafer using standard  shadow mask e-beam lithography techniques with two-angle evaporation of Aluminum. Each of the four arms contains an overlay tunnel junction of area $100 \times 100$ nm$^2$.

The sample is slightly asymmetric in the four arms, with resistances given in Table \ref{tunnelR}. The Fano factor of a separate single tunnel contact was measured and $F = 1$ was observed. Hence, our tunnel junctions provide excellent classical shot noise generators. The thickness of Al was $20-40$ nm and the diffusion constant was estimated to be $D=\frac{1}{2} v_F \ell \sim 80$ ${\rm cm^2}/s$ where $v_F=1.6 \times 10^6$ m/s and $\ell  \simeq 8$ nm are the Fermi velocity and elastic mean free path of electrons, respectively. The mean free path was calculated from the measured resistivity of Aluminum at 4.2 K. Because of the small volume, the electron phonon coupling is not sufficient to thermalize the electrons, and the distribution will become non-equilibrium \cite{Giazotto2006}.  Depending on the electron-electron scattering rates, the distribution on the island may be either a two-step distribution \cite{Pothier1997}, for negligible scattering, as assumed in our theoretical model, or a quasi-equilibrium state \cite{Steinbach1996} for stronger scattering.

\begin{table}
\begin{center}
\begin{tabular}{llll}
\hline\noalign{\smallskip}
$R_1$ & $R_2$ & $R_3$ &  $R_4$ \\
49 k$\Omega$ & 53 k$\Omega$ & 37 k$\Omega$ & 40 k$\Omega$\\

\noalign{\smallskip}\hline
\end{tabular}
\caption{Parameters of the sample. The resistance values were determined at 4.2 K.
 }
\label{tunnelR}
\end{center}
\end{table}

%

\begin{figure}[!t]
\begin{center}
\includegraphics[width=0.85\linewidth]{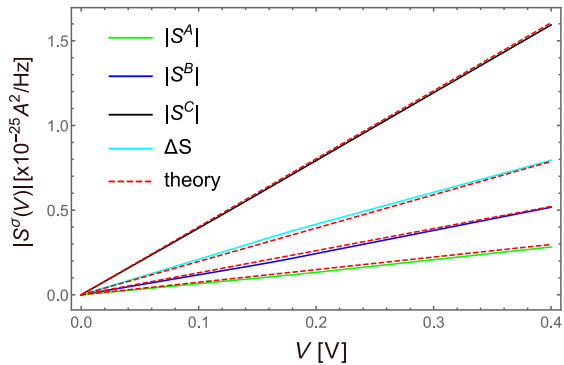}
\caption{(Color online.) Cross-correlation data obtained for the four-terminal tunnel junction biased in configurations $A$, $B$, $C$, and the corresponding HBT effect $\Delta S$. Theoretical curves match within few $\%$ the measurement. The fitting parameter amounts to rescaling the $y$-axis, accounting for a scaling of $\beta=0.9$ compared to the theoretical values obtained using Eq.~\ref{fitnoise}.  }
\label{tunneljunction}
\end{center}
\end{figure}

Cross-correlation expressions for our junction are obtained from Eq.~\ref{totalnoise} by setting $F_j=1$,
\begin{align}
S^C=&2eV\frac{G_2G_4}{G^2_{\Sigma}}\frac{\left(G_1+G_3\right)^2}{G_{\Sigma}},\label{fitnoise}\\
S^A=&S^C \frac{G_1^2}{\left(G_1+G_3\right)^2},\quad S^B=S^C \frac{G_3^2}{\left(G_1+G_3\right)^2}.\notag
\end{align}
For symmetric arm resistances, $S^C/S^{A,B}=4$ follows from Eqs.~\ref{langeneqn} and \ref{fitnoise}.
The asymmetry in the measured sample leads to modified cross-correlation ratios, e.g. $S^C/S^A$ increases to $5.6$ in the experiment. Using Eq.~\ref{fitnoise} we find $(S^C/S^A)_\text{theory}=(1+G_3/G_1)^2=5.40$, within $4\%$ of the measured value. The measured ratio $S^C/S^B$ is $3.1$, matching the theoretical value $(S^C/S^B)_\text{theory}=(1+G_1/G_3)^2=3.08$ within $1\%$. The ratio $S^C/(S^A+S^B)=(G_1+G_3)^2/(G_1^2+G_3^2)$ is maximal for identical contacts \cite{Langen1997}, where it reaches the value $2$. For our asymmetric sample the measured value is $2.0$ with the theoretical result given by $1.96$. The characteristic HBT ratio is $\Delta S/ (S^A+S^B)=1.0$, in close agreement with the theoretical value. 

The measured data is depicted as a function of bias voltage in Fig. \ref{tunneljunction}. The theoretical calculations are based on Eq.~\ref{fitnoise}, including a dimensionless scaling used as fitting parameter. The dimensional scaling is given by $\beta=S_\text{exp}/S_\text{theory}=0.9$, accounting for a small difference between the calculated value and the calibration of the measurement scheme.

The measurement has been performed for a wide range of voltage bias. At zero bias, the measured cross-correlation was negligible. At large bias, the I-V curve becomes non-linear giving rise to the weak non-linearity observed in Fig. \ref{tunneljunction}. Good agreement between our theoretical model and the measurement suggests that even for large applied voltage, inelastic processes in the junction remain weak, insufficient to relax the distribution function of the metallic island to its equilibrium value.

\section{Conclusion}
\label{sec:concl}
We have measured the Hanbury--Brown and Twiss exchange correlations in an asymmetric four-terminal tunnel junction using three biasing configurations. We demonstrate that although contacts are classical shot noise generators, the current cross-correlations are large, reaching close to their theoretical maximum. Our measurements agree closely to predictions of a simple theoretical model where cross-correlations arise due to fluctuations of the electric potential of the metallic island. The current experiment and theoretical framework provide an important benchmark for future investigations of noise in junctions with higher transparency contacts. For such contacts, our model predicts that quantum shot noise characterized by Fano factor $F<1$ gives rise to a reduction of the HBT signal $\Delta S$ from the value $(S^A+S^B)$ measured here. The HBT ratio $\Delta S/(S^A+S^B)$ may be significantly reduced and can become negative for shot noise characterized by Fano factor $F<1/3$.

\begin{acknowledgements}
We acknowledge fruitful discussions with Ya. Blanter, T. Elo, C. Flindt, G. Lesovik, T. Heikkil\"a, P. Virtanen and M. Wiesner.
Our work was supported by the Academy of Finland (contract 250280, Centre of Excellence LTQ), Finnish National Doctoral Programme in Nanoscience (NGS-NANO) and National Doctoral Programme in Materials Physics (NGSMP). This research project made use of the Aalto University OtaNano/LTL infrastructure.
\end{acknowledgements}

\end{document}